\documentclass[twocolumn, aps]{revtex4}
\usepackage{graphicx}
\usepackage[intlimits]{amsmath}

\begin{document}

\author{Michael L. Gorodetsky}
\affiliation{ Faculty of Physics, M.V. Lomonosov Moscow State University}

\title{Thermal noises and noise compensation in high-reflection multilayer coating}

\begin{abstract}
Thermal fluctuations of different origin in the substrate and in the coating of optical mirrors produce phase noise in the reflected wave. This noise determines the ultimate stabilization capability of high-Q cavities used as a reference system. In particular this noise is significant in interferometric laser gravitational wave antennas. It is shown that simple alteration of a mirror multilayer coating may provide suppression of phase noise produced by thermorefractive, thermoelastic, photothermal and thermoradiation induced fluctuations in the coating.
\end{abstract}

\maketitle

\section{Introduction}

In different optical systems phase noise and shot noise are usually considered as the main fundamental limiting factors of sensitivity. These noises determine the standard quantum limit of classical gravitational wave antennas \cite{noisegw}. At the same time in laser systems coupled with high-finesse optical resonators fundamental frequency stability may be determined also by other fundamental effects connected with mechanical, thermodynamical and quantum \cite{bragstab} properties of solid boundaries. Many of these effects firstly identified and calculated on the forefront of laser gravitational wave antennas \cite{Levin98, BGV99, BGV00, BV03, Fejer, Harry} are becoming increasingly important in other high-Q optical systems  \cite{Numata,Webster,Microsphere,Matsko1,Matsko2}.

\section{A noisy zoo}

Excess optical phase noise is added to the wave reflected from the mirror of the optical cavity due to variation of boundary conditions produced by fluctuations of the surface and refractive index in the surface layer. 

{\bf I. Substrate Brownian noise}. Historically the first noise of this origin identified as a problem in gravitational wave antennas was thermal intrinsic noise produced by internal friction which may be related to the surface using fluctuation-dissipation theorem \cite{Levin98}. This noise is frequently noted as ``Brownian motion'' or simply thermal noise.

The spectral density of surface fluctuations of a mirror averaged over the Gaussian beam radius $w$ (determined at $1/e^2$ decay of intensity) is the following:   
\begin{eqnarray}
  S_{B}^{sub}=\frac{2 k_B T \phi(f) (1-\sigma_s^2)}{\pi^{3/2} Y_s w f}, \label{SDbulk}
\end{eqnarray}
where $k_B$ is the Boltzmann constant, $T$ is the temperature of the mirror, $\phi(f)$ is the loss angle of the mirror substrate at a frequency $f$ of analysis, $\sigma_s$ is the Poisson ratio and $Y_s$ is the Young's modulus of the substrate.

{\bf II. Coating Brownian noise}. In the same way internal friction in the material of the coating produces additional fluctuations, which may be significantly stronger due to high level of acoustical losses in the material of the layers even though the thickness of this coating is relatively small. Effective thermal noise associated with the coating is determined by the same relation (\ref{SDbulk}), with \cite{Harry,Titana}:

\begin{eqnarray}
&&S_{B}^{coat}=\frac{2 k_B T \phi_{coat}(f) (1-\sigma^2)}{\pi^{3/2} Y_s w f}, \label{SDcoat}\\
&&\phi_c\simeq \frac{d_N(1-2\sigma_s)}{\sqrt{\pi} w(1-\sigma_s)(1-\sigma_\Vert)}\times \nonumber\\
&&\times\Bigl[\frac{Y_\bot(1-\sigma_\Vert)-2\sigma^2_\bot Y_\Vert}{Y_\bot^2(1+\sigma_s)(1-2\sigma_s)}Y_s\phi_\bot
+\frac{Y_\Vert\sigma_\bot(\phi_\Vert-\phi_\bot)}{Y_\bot}  \nonumber\\
  &&+ \frac{Y_\Vert(1+\sigma_s)(1-2\sigma_s)\phi_\Vert}{Y_s(1+\sigma_\Vert)}\Bigr]\nonumber\\
&&\phi_\bot=Y_\bot\langle \phi/Y\rangle \;\;\;\;\;
  \phi_\Vert=\langle Y\phi\rangle/Y_\Vert \;\;\;\;\;
Y_\bot=\langle 1/Y\rangle^{-1}\;\;\;\;\;\nonumber\\
  &&Y_\Vert=\langle Y\rangle\;\;\;\;\;\sigma_\bot=\langle \sigma Y\rangle/\langle Y\rangle
  \;\;\;\;\;\sigma_\Vert\simeq(\sigma_l+\sigma_h)/2\nonumber
\end{eqnarray}
Here and below averaging over the multilayer coating is performed as follows:
\begin{eqnarray}
\langle X\rangle \equiv \frac{X_ld_l+X_hd_h}{d_l+d_h}\label{averaging}.
\end{eqnarray}
Indexes $l$ and $h$ denote constants for low and high reflection coating layers with thicknesses $d_l$ and $d_h$, and $d_N=N(d_l+d_h)$ is the total thickness of the coating with $N$ double layers. This noise has been experimentally measured in specialy designed Thermal Noise Interferometers (TNI) and may limit the sensitivity of the next generation of gravitational wave antennas \cite{Titana}. 

{\bf III. Substrate thermoelastic noise}. In \cite{BGV99} Braginsky, Gorodetsky and Vyatchanin (BGV) suggested that temperature fluctuations  in the bulk of the mirror transformed through thermal expansion to surface fluctuations should produce additional mechanical noise. Using both the FDT approach and the Langevin method for the analysis of thermal correlation functions analogous to that developed earlier by van Vliet  \cite{vanVliet, Vlietfilm} they have shown that this noise may be understood as the noise produced by thermoelastic damping. 
\begin{eqnarray}
  S^{sub}_{TE}=\frac {4k_B T^2 \alpha_s^2 (1+\sigma_s)^2 \kappa_s}{\pi^{5/2}(C_s\rho_s)^2 w^3f^2},
\end{eqnarray}
where $\alpha_s$ is the coefficient of thermal expansion, $C_s$ is specific heat capacity, $\kappa_s$ is thermal conductivity, and $\rho_s$ is the density of the substrate.

{\bf IV. Coating thermoelastic noise}. The same as III but noise is caused by thermal expansion of the coating or from another point of view by coating thermoelastic losses \cite{BV03,Fejer}:

\begin{eqnarray}
&&S^{coat}_{TE}=\frac{8k_B T^2 (1+\sigma_s)^2\alpha_c^2d^2_N}{\pi^{3/2} \sqrt{\kappa_s C_s\rho_s} w^2 f^{1/2}}G(\omega)_{TE}^{coat}\\
&&\alpha_c= \nonumber\\
&&\frac{1}{2} \left\langle \frac{\alpha}{1-\sigma}\left(\frac{1+\sigma}{1+\sigma_s}+(1-2\sigma_s)\frac{Y}{Y_s}\right)\right\rangle
- \alpha_s \frac{\langle C\rho\rangle}{C_s \rho_s}\nonumber\\
&&G(\omega)_{TE}^{coat} = \nonumber\\
&&\frac{2}{R\xi^2} \frac{\sinh \xi - \sin \xi + R(\cosh \xi - \cos \xi)}{\cosh\xi +\cos\xi + 2R\sinh\xi+R^2(\cosh\xi-\cos\xi)}\nonumber\\
&&\xi=\sqrt{2\omega d_N^2\langle\rho C\rangle \langle\kappa^{-1}\rangle} \quad\quad R=\sqrt{\frac{\langle\rho C\rangle}{C_s\rho_s\kappa_s\langle\kappa^{-1}\rangle}}\nonumber
\end{eqnarray} 

For silica-tantala coating $\alpha_c$ is within 10\% from simpler $\langle\alpha\rangle$ averaged as (\ref{averaging}) and for $\xi\ll 1$  the following approximation is valid:
\begin{eqnarray}
G(\omega)_{TE}^{coat}\simeq 1-\frac{3R^2-1}{3R}\xi \label{gappr}
\end{eqnarray}
 
{\bf V.  Coating thermorefractive noise}. In \cite{BGV00} BGV identified another effect of thermal fluctuations in a mirror -- due to thermorefractive factor $\beta_{l,h}=dn_{l,h}/dT$ in the layers of the coating, the phase of the reflected way should be also proportional to these thermal fluctuations: 
\begin{eqnarray}
S^{coat}_{TR}= \frac{2 k_B T^2 \beta^2_{eff}\lambda^2}{\pi^{3/2} \sqrt{\kappa_s\rho_s C_s}w^2 f^{1/2}}\nonumber\\
\beta_{eff}=\frac{1}{4}\frac{\beta_hn_l^2+\beta_ln_h^2}{n_h^2-n_l^2}
\end{eqnarray}
This value of $\beta_{eff}$ is valid only if the topmost covering quarter-wavelength layer (cap) has lower refractive index $n_l<n_h$ and thickness $d_c=\lambda/(4n_l)$ ($\lambda$ is the wavelength) and incorrect in the opposite case, when $d_c=0$. More general case for an arbitrary $d_c$ is analyzed in Appendix A and discussed below. The existence of excess thermorefractive phase noise was earlier predicted \cite{Wanser} and measured \cite{Knudsen} in optical fibers. This noise was also measured in high-Q optical microspheres \cite{Microsphere}.

{\bf VI.  Substrate photothermoelastic noise}. Noise may be produced not only by intrinsic thermal fluctuations but also by fluctuation of absorbed power, heating the mirror \cite{BGV99}: 
\begin{eqnarray}
S^{sub}_{PTE}= \frac{\alpha^2 S_{abs}}{\pi^4 \rho_s^2 C_s^2 w^4 f^2},
\end{eqnarray}
where $\omega_0$ is optical frequency and $S_{abs}$ is spectral density of the absorbed power. For the case of shot noise in the absorbed power $P_{abs}$, single sided spectral density $S_{abs}=2\hbar\omega P_{abs}$. More general case of arbitrary power fluctuations was analyzed in \cite{Shi-Chen}.

{\bf VII, VIII  Coating photothermoelastic and photothermorefractive noises}. Analogously to IV and V extrinsic thermal fluctuations produced by absorbed optical power should produce coating noise due to thermal expansion and fluctuations of the refractive index in the surface layers, with approximate estimates obtained in \cite{Rao}, which are verified in Appendix B using BGV approach, where low frequency approximations for $G^{coat}_{surf}(\omega)$ are also obtained:
\begin{eqnarray}
S^{coat}_{PTE}= \frac{4S_{abs} (1+\sigma_s)^2 \alpha_c^2 d^2_N}{\pi^3 \rho_s C_s\kappa_s w^4 f} G^{coat}_{surf}(\omega)
\end{eqnarray}
\begin{eqnarray}
S^{coat}_{PTR}= \frac{S_{abs} \beta^2_{eff} \lambda^2}{\pi^3 \rho_s C_s\kappa_s w^4 f} G^{coat}_{surf}(\omega)
\end{eqnarray}

{\bf IX, X, XI Substrate and coating Stefan-Boltzmann thermoradiation noises}.
In all the above derivations Stefan-Boltzmann radiation from the surface was neglected considering that thermal conductivity dominates this process. However thermal radiation as a dissipative process produces additional fluctuations of temperature applied to the mirror. The spectral density of these temperature fluctuations obtained by van Vliet is given in \cite{vanVliet, Vlietfilm}. As this noise is applied to the surface practically in the same way as photothermal noise, we can use expressions obtained in VI, VII, VIII to estimate thermoradiation elastic noise in the substrate and in the coating and thermoradiation refractive noise by simple substitutions 
\begin{eqnarray}
S_{abs}\to S_{SB}=8\sigma_B k_BT^5\pi w^2
\end{eqnarray}

All the equations for the noises above are given or normalized for the case of infinite semi-space and adiabatic frequency range and show the ceiling of the noises in the range $\frac{D_c}{d_N^2}\gg f\gg\frac{D_s}{w^2}$ ($D=\kappa/(\rho C)$ is thermal diffusivity). This range is mostly valid for the operating frequencies of LIGO. Corrections for higher and lower frequencies and finite size mirrors where needed, which could be essential for other applications with methods of accounting them for some cases may be found in \cite{Liu} and \cite{Cerdonio}. 

\begin{figure}
\centerline{\includegraphics[width=0.45\textwidth]{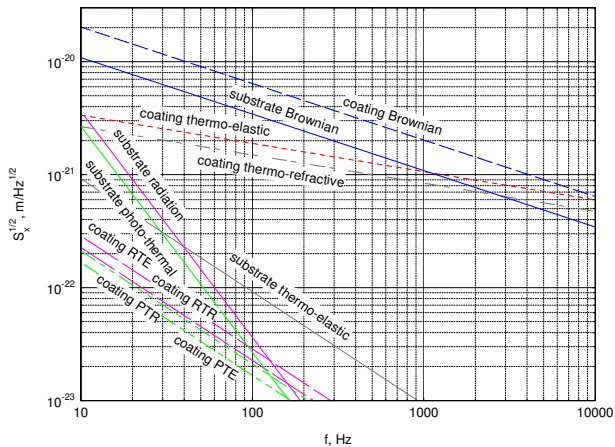}}
\caption{Thermodynamic noises in the LIGO mirrors}
\label{fig1}
\end{figure}

All the noises numbered above are plotted on Figure 1 for the parameters of LIGO mirrors given in Appendix E. It is clearly seen that currently only Brownian and coating thermoelastic and thermorefractive noises are essential. It is appropriate to note however that the Brownian noises are determined by the loss factor which is gradually improving with the development of technology and has not reached as it looks a fundamental shelf. This is especially true for the case of coating Brownian noises where the loss factors in thin films are several order higher than in the bulk. It is quite possible that upcoming progress in the technology of dielectric film deposition can significantly improve the situation. It is worth to note in particularly recent significant improvement of tantala layers losses by addmixing of titanium oxide \cite{Titana}. Another improvement is possible by using multilayer coating consisting with unequal layers and in particular generalized quarter-length layers where high-refractive layer has smaller and low refractive has larger optical thickness, preserving half a wavelength in the sum \cite{Ovcharenko}. As it is high-reflective material that has significantly higher losses and larger thermal expansion it becomes possible to lower Brownian surface and thermoelastic noises optimizing relative thickness and compensating decreased reflectivity by additional layers \cite{Castaldi}. 

Thermorefractive and thermoelastic noises unlike Brownian noises are determined by basic constants of the material which could not be easily modified. These noises can become dominating at frequencies higher than 1 KHz in projected LIGO mirrors. One may notice that these two noises having the same order of magnitude and the same dependence from frequency and beam waist are produced by the same mechanism -- temperature fluctuations as well as other pairs VII-VIII and X-XI where these fluctuations of temperature are produced by external source or by surface radiation. Surprisingly a natural question about correlation of these noises has not been explored so far.

\section{Compensation of thermo-optical noises}

H.J.Kimble recently proposed a magnificent idea \cite{Kimble}, that surface fluctuations and refractive index change  produced  by strain could compensate each other in specially designed multilayer coating. In principle these strain produced fluctuations should be added to the ``noisy zoo'', however strain fluctuations should be significantly suppressed near the free surface. As Brownian fluctuations are currently the main limiting factor in LIGO mirrors, the realization of this idea can radically improve the sensitivity. 

It is shown below that the same idea can be very easily applied to the surface noises produced by temperature fluctuations, namely to thermoelastic (IV) + thermorefractive (V) \cite{EvansHarry} as well as pairs of photothermal (VI+VII) and  thermoradiation (X+XI) noises.

Phase fluctuations produced by thermal refraction are equivalent to those produced by effective surface fluctuations $\delta x_{eff} =\delta\phi = \lambda \beta_{eff}\delta T$. I have removed a confusing minus in the definition of $\beta_{eff}$ as compared to BGV \cite{BGV00} to have $\beta_{eff}$ positive and have the same sign as $\beta = \frac{dn}{dT}$, which is positive in most of optical materials). The effect of thermal refraction leads to lengthening of optical thickness $n d$ and moves the effective surface from which the beam is reflected deeper in the mirror away in the direction of incoming beam. At the same time thermal expansion moves the surface of the mirror in the opposite direction toward the beam. This speculation and an observation that the value of thermorefractive and thermoelastic coating noises are close enough (Fig.1) gives hope that these two effects can compensate each other if some parameters of the coating are tweaked.

\begin{figure}
\center\includegraphics*[width=0.45\textwidth]{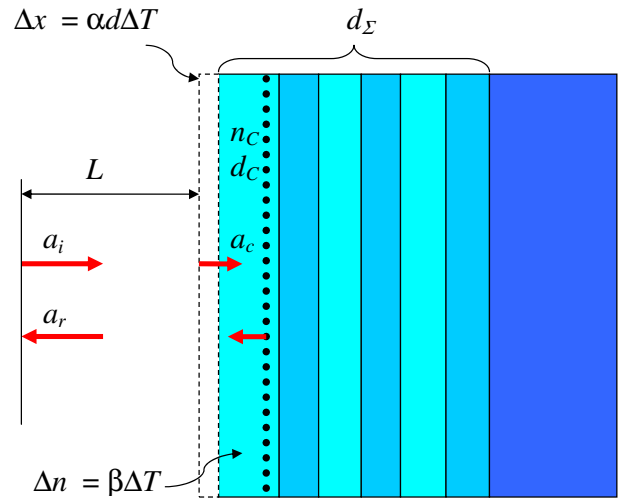}
\caption{Schematic of thermal noise compensation}
\label{fig2}
\end{figure}

The simplest model (Fig.2) can illustrate the effect of thermal noise compensation. Let we have $2N$ layers deposited on unmovable substrate and let the wave is reflected from the outermost layer only. The system of equations for the stationary amplitudes of circulating and reflected waves
\begin{eqnarray}
a_c&=&a_i \sqrt{1-r^2} e^{ikL} + r a_c e^{2ikn_c d_c}\nonumber\\
a_r&=&-ra_0e^{2ikL}+\sqrt{1-r^2} a_0e^{2ikn_cd_c+ikL}
\end{eqnarray}  
I use $\propto e^{-i\omega t+ikx}$ for a plain wave incident on the mirror, where $k=2\pi/\lambda$ is the wave number. With the reflection coefficient $r=\frac{n_c-1}{n_c+1}$ the system has the following solution:
\begin{eqnarray}
\Gamma_0 &=& \frac{a_r}{a_i}=e^{2ikL}\frac{e^{i\phi_c}-r}{1-re^{i\phi_c}}\nonumber\\
\phi_c&=&2kd_cn_c
\end{eqnarray}  
Now if due to the change of temperature the index of refraction and the length are changed $n_c\to n_c+\beta \Delta T$, $d_cn_c\to d_cn_c(1+\beta/n_c\Delta T+\alpha \Delta T)$, and $L\to L-\alpha\, d_N\,\Delta T$,
\begin{eqnarray}
\Gamma_0 &\to& \Gamma_0 e^{i\Delta\phi}\\
\Delta\phi&=&-\left[\alpha d_N-2n_cd_c\frac{\beta(1+\mbox{ sinc}\,\phi_c)+n_c\alpha}{n^2+1-(n^2-1)\sin\phi_c}\right]2k\Delta T\nonumber
\end{eqnarray}  

One sees that though the last term in brackets is oscillating, its average linear dependence on $d_c$ allows to compensate the first term caused by thermal expansion by the thermal dependence of the refraction index with appropriate choice of  the cover layer thickness.

In Appendix A it is shown that the same situation is also valid for multilayer coating with varying outermost low-refractive layer. 
\begin{eqnarray}
&&\beta_{eff,N\to\infty}=\\
&&\frac{1}{2\pi}\frac{\pi n_l^2(\beta_l+\beta_h)+ \beta_l (\phi_c- \sin\phi_c)(n_h^2-n_l^2)}{(n_h^2-n_l^2)(n_l^2+1+(n_l^2-1)\cos\phi_c)}, \label{infinite0}\nonumber
\end{eqnarray}
The same type of expressions for limited number of layers and unequal layers are too bulky for practical purposes and it is easier to calculate them numerically by recurrent procedures.

\begin{figure}
\center\includegraphics*[width=0.45\textwidth]{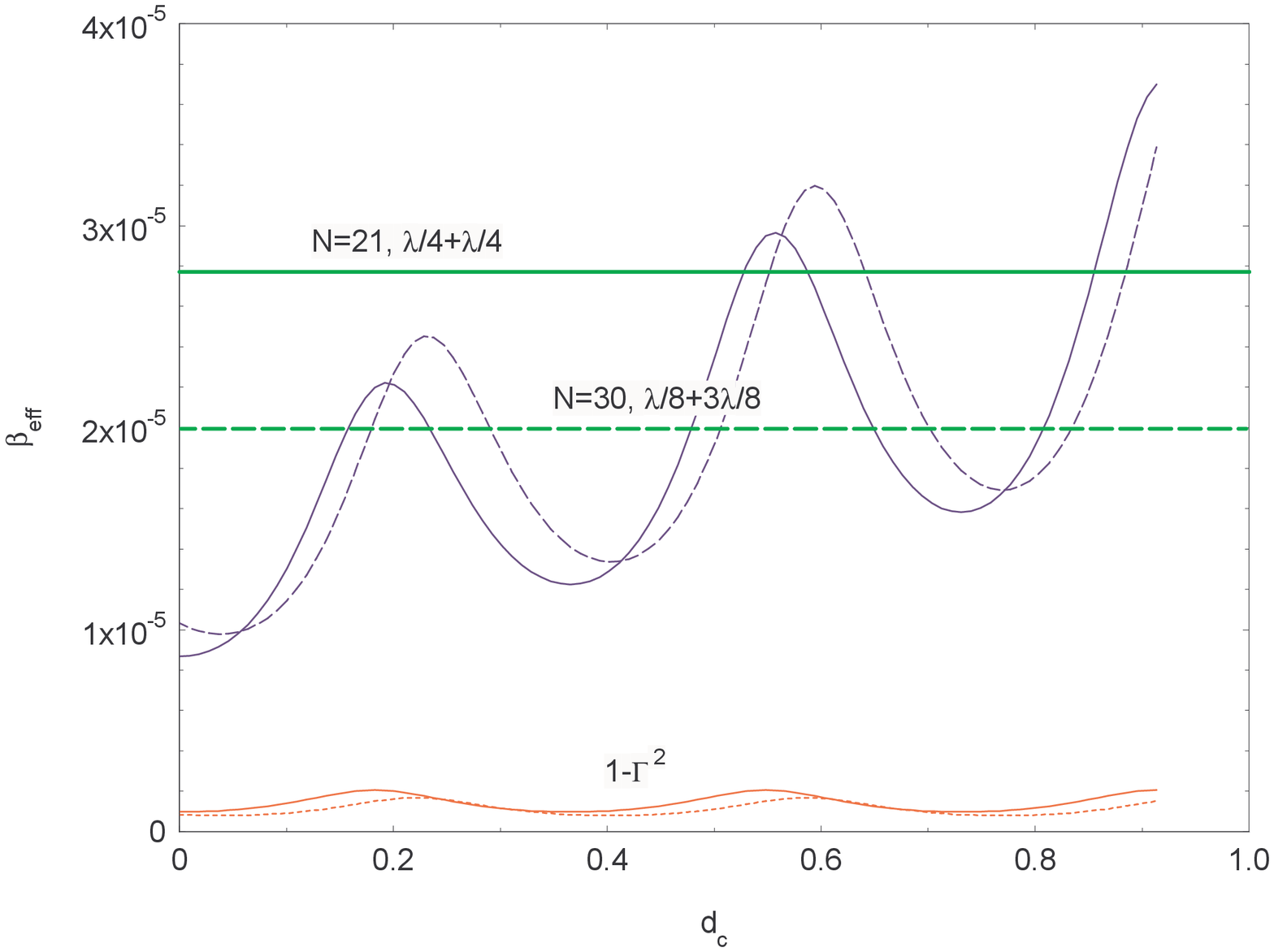}
\caption{Leveling of thermorefractive and thermoelastic constants}
\label{fig3}
\end{figure}

Fig.\ref{fig3} plots the dependence of absolute value of $\beta_{eff}$ from the thickness of the covering layer. Horizontal lines show the equivalent level of thermoelastic parameter $2\alpha_c(1+\sigma_s)/\lambda$. The crossings of the $\beta$-curve with horizontal $\alpha$-lines provide the points of compensation (at low frequencies). Solid curves are calculated for the QWL coating with 21 layer, while dashed curves stand for optimized unequal 30 double layers with $d_h=\lambda/8$ and $d_l=3\lambda/8$, providing approximately the same low transmittivity $\sim 10^{-6}$. The dependence of transmittivity (neglecting optical losses) on the thickness of the topmost layer is shown in the bottom of Fig.\ref{fig3}. In most coating designs they choose the cap to have zero zero or $\lambda/2n_l$ (protective cap) thickness giving the best reflectivity, i.e. minimal transmittivity. However degradation of reflectivity can be more than compensated by one additional pair of deep layers if the noise issues are considered as more significant. It is possible to tweak simultaneously both high- and low-reflective top layers. As thermorefractive noise at $d_c=0$ is somewhat smaller than thermoelastic noise in a wide range of frequencies both for equal QWL and for layers with optimized thicknesses, noise compensation benefits from optimized levels in the same way as thermoelastic noise. 

Noise cancellation can not be absolute. The effective volumes of thermorefractive and thermoelastic layers are significantly different as the wave is mostly reflected from a few outermost layers where the optical power decays exponentially and the thermal expansion is provided by the whole structure of the coating. The analysis of the level of possible compensation is given in Appendix C. 

\begin{figure}
\center\includegraphics*[width=0.45\textwidth]{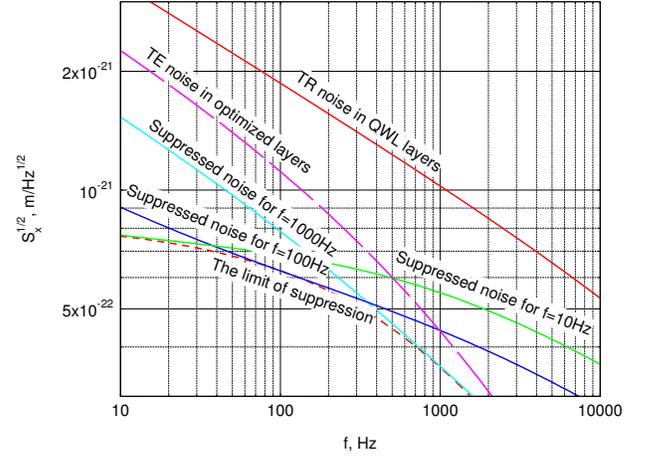}
\caption{Compensation of thermorefractive and thermoelastic noises in the coating as optimized for different frequencies}
\label{fig4}
\end{figure}

The results of this analysis are shown on Fig.\ref{fig4} for the case of optimized layers ($d_h=\lambda/(8n_h)$, $d_l=3\lambda/(8n_l)$) and cover layer optimized for the frequencies of operation 10, 100 and 1000 Hz. For comparison thermorefractive noise in QWL layers leveled at thermoelastic noise is plotted. One can see that the proposed simple idea of suppression of thermorefractive and thermoelastic noises allows to lower their combined level up to 3 times in the vicinity of 100Hz-1KHz range. If the noise is compensated for zero frequency, the combined noise becomes frequency independent at low frequencies. At high frequencies, when the length of thermal diffusion becomes comparable with the thickness of the coating, the spectral density of thermoelastic noise falls more rapidly than that of thermorefractive noise and compensation though at lower level becomes worse. Roughly speaking, simple leveling of $alpha$ and $beta$ parameters should work for low frequencies below 100 Hz and QWL covering layer is close to optimal, while for higher frequencies very thin or half-wavelength covering layer, providing minimal $\beta_{eff}$ is better. It is appropriate to note however, that the method of analysis used in Appendix D becomes too approximate in this case and more accurate calculations are required for this case, which are forthcoming. 

It is instructive to look also whether it is possible to compensate with thermorefractive effect not only surface but substrate thermoelastic noise. The analysis is given in Appendix D. It is shown that because of quite different effective working volumes of these effects the level of compensation is negligibly small for LIGO mirrors ($\sim\sqrt{D_s/(w^2f_{opt})}\simeq 0.015/\sqrt{f_{opt}}$), moreover in LIGO the combined even compensated effect in the coating is still higher than the effect in the bulk for the operating frequencies, however this compensation may be essential in more traditional interferometers with significantly smaller beam waists.

\section*{Conclusion} 

Summing up the results of the presented analysis we see that the list of essential noises in high-reflection mirrors by simple tweaking of a multilayer coating may be noticeably shortened and we return to the situation of a decade ago when only Brownian noise was considered.  

\begin{acknowledgments}
I am grateful to V.B.Braginsky and S.P.Vyatchanin for stimulating and encouraging discussions.  
\end{acknowledgments}

\appendix

\section*{Appendix A. Multilayer coating}

\begin{figure}
\center\includegraphics*[width=0.45\textwidth]{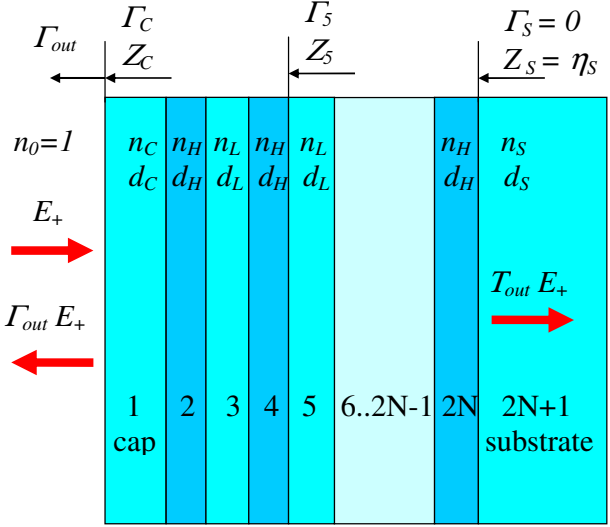}
\caption{Multilayer coating}
\label{fig5}
\end{figure}

Let we have a multilayer coating, consisting of $2N$ alternating layers (or N double layers) on a substrate, with 
the first being a cover layer and the substrate denoted by $2N+1$ (Fig.\ref{fig5}).

The method of calculation of properties of a multilayer reflection coating is quite simple and computationally straightforward (see for example \cite{Haus}), however not without pitfalls. Though several analytical expressions are obtained below, for different practical applications it is usually more appropriate to do numerical anaysis. 

Taking the dependence of forward and backward plane waves $E\!,\!H_{\pm}(z)$ as $e^{-i\omega t\pm ikz}$, $E\!,\!H(z)=E\!,\!H_+(z) + E\!,\!H_-(z)$ one introduces two values -- amplitude coefficient of reflection $\Gamma(z) = E_{-}(z)/E_{+}(z)$ and impedance $Z(z)= E(z)/H(z)$, connected by the following equations:
\begin{eqnarray}
Z(z)&=&\frac{E(z)}{H(z)}={\eta}\frac{1+\Gamma}{1-\Gamma}\nonumber\\
\Gamma(z)&=&\frac{E^-(z)}{E^+(z)}=\frac{Z-\eta}{Z+\eta}\nonumber\\
\eta&=&\sqrt{\frac{\mu\mu_0}{\epsilon\epsilon_0}}=\frac{1}{n}\sqrt{\frac{\mu_0}{\epsilon_0}} = \frac{1}{n}Z_0
\end{eqnarray}
As tangential components of electrical and magnetic fields are continuous on the boundary, the same is true for the impedance $Z$. From the other hand, $\Gamma$ jumps on boundaries but change phase only in homogeneous media $\Gamma(z+d)=\Gamma(z)e^{-i2knd}=\Gamma(z)e^{-i\phi}$. In this way, starting from the rightmost layer, where in the substrate $\Gamma_s=0$ and $Z_s=\eta_s$, moving layer by layer from right to left, we can calculate $\Gamma_{out}$.
\begin{eqnarray}
&&\Gamma_s=\Gamma_{2N+1}=0\nonumber\\
&&\Gamma_{j}=\frac{Z_{j+1}-\eta_j}{Z_{j+1}+\eta_j}e^{i\phi_j}=\frac{\Gamma_{j,j+1}+\Gamma_{j+1}}{1+\Gamma_{j,j+1}\Gamma_j}e^{i\phi_j}\nonumber\\
&&\Gamma_{out}=\frac{\Gamma_{0C}+\Gamma_c}{1+\Gamma_{0C}\Gamma_c}\nonumber\\
&&\Gamma_{j,j+1}=\frac{\eta_{j+1}-\eta_{j}}{\eta_{j+1}+\eta_{j}}=\frac{n_{j}-n_{j+1}}{n_{j}+n_{j+1}}
\end{eqnarray}

The most frequently used type of multilayer coating is the one consisting of alternating high and low refractive quarter wave layers (QWLs) with $n_{2j}=n_h$, $n_{2j+1}=n_l$, $d_j=\frac{\lambda}{4n_j}$, $\phi_j=2nkd_j=\pi$. This type of coating produces the highest reflectivity for the given number of layers. It is easy to show that in this case $Z_j=\frac{\eta_j^2}{Z_{j+1}}$. For two different cases 1) when the cover layer $n_c=n_l$ is also a QWL -- low refraction output layer, and without the cover layer ($d_c=0$, high refractive output layer) one gets:
\begin{eqnarray}
\Gamma_{out,l}&=&\frac{1-\varepsilon n_s}{1+\varepsilon n_s}\nonumber\\
\Gamma_{out,H}&=&-\frac{1-\varepsilon n_s/n_l^2}{1+\varepsilon n_s/n_l^2}, \label{finite}
\end{eqnarray}
where $\varepsilon=(n_l/n_h)^{2N}$. It is essential that when $N\to\infty$ and $n_l/n_h<1$ both cases give $|\Gamma_{out}|=1$ but with different signs.

When due to some homogeneous fluctuation in the mirror coating thicknesses and refractive index change, the output phase will also change. To calculate this phase fluctuation in a general case of arbitrary cover layer we use the model of infinite coating \cite{BGV00}.

Let we have a multilayer coating with infinite number of alternating
layers of optical thickness $n_ld_l$ and $n_hd_h$ and in some layer
$n_h$ near its left boundary we have the reflectivity $\Gamma_\infty$. Now we
add from the left two other layers $n_l$ and $n_h$.

\begin{eqnarray}
&&\Gamma_l=\frac{\Gamma_{lh}+\Gamma_\infty}{1+\Gamma_{lh}\Gamma_\infty}e^{2ikd_ln_l}=\frac{\Gamma_{lh}+\Gamma_\infty}{1+\Gamma_{lh}\Gamma_\infty}e^{i\phi_l}\nonumber\\
&&\Gamma_h=\frac{\Gamma_{hl}+\Gamma_l}{1+\Gamma_{hl}\Gamma_l}e^{2ikd_hn_h}=\frac{\Gamma_l-\Gamma_{lh}}{1-\Gamma_{lh}\Gamma_l}e^{i\phi_h}
\end{eqnarray}

As again we have the infinite number of the same type layers then $\Gamma_h=\Gamma_\infty$ and we can solve the system for the $\Gamma_\infty$: 
\begin{eqnarray}
&&\Gamma^2_\infty +\frac{1-e^{i\phi_l}e^{i\phi_h}+\Gamma^2_{lh}(e^{i\phi_h}-e^{i\phi_l})}{\Gamma_{lh}(1-e^{i\phi_l})}\Gamma_\infty+e^{i\phi_h} =0
\label{Gamma}
\end{eqnarray}

We may find the solution of this equation for the generalized QWL layers with $2 k n_{l,h} d_{l,h}= \pi\pm \psi_g$, $e^{i\phi_{l,h}}=-e^{\pm i\psi_g}$. 

Substituting $\Gamma_\infty=e^{i\phi_\infty}$ in (\ref{Gamma}), we obtain the following equation:
\begin{eqnarray}
\sin\phi_\infty+\sin(\phi_\infty+\psi_g)+\Gamma_{lh}\sin\psi_g=0,
\end{eqnarray}
and after some manipulations we find:
\begin{eqnarray}
\phi^l_\infty&=&-\frac{\psi_g}{2}-\arcsin\left(\Gamma_{lh}\sin\frac{\psi_g}{2}\right)\nonumber\\
\phi^H_\infty&=&-\frac{\psi_g}{2}+\arcsin\left(\Gamma_{lh}\sin\frac{\psi_g}{2}\right)+\pi
\end{eqnarray}

\begin{eqnarray}
n_l\to n_l+\beta_l u\nonumber\\
n_h\to n_h+\beta_h u 
\end{eqnarray}

\begin{eqnarray}
&&\Gamma^h_\infty(u)=e^{i\phi_{\infty,h}}(1+\gamma_h u)\nonumber\\
&&\gamma_h = -\frac{i}{4\sqrt{1-\Gamma^2_{lh}\sin^2\frac{\psi_g}{2}} n_ln_h\Gamma_{lh}\cos\frac{\psi_g}{2}}\times\nonumber\\
&&\biggl[n_l\beta_h(\pi-\psi_g)\Bigl(\Gamma_{lh}\cos\frac{\psi_g}{2}-\sqrt{1-\Gamma^2_{lh}\sin^2\frac{\psi_g}{2}}\Bigr)^2\nonumber\\
&&+n_h\beta_l(\pi+\psi_g)(1-\Gamma_{lh}^2)\biggr]
\end{eqnarray}

Now we add a cover layer with $d_c$ and $n_c=n_l$ and calculate $\Gamma_{out}$:
\begin{eqnarray}
\Gamma_c&=&\frac{n_l-n_h+(n_l+n_h)\Gamma^h_\infty}{n_l+n_h+(n_l-n_h)\Gamma^h_\infty}e^{2ikd_cn_l}\nonumber\\
\Gamma_{out}&=&\frac{1-n_l+(1+n_l)\Gamma_c}{1+n_l+(1-n_l)\Gamma_c}
\end{eqnarray}

In particular case of vanilla QWL layers under homogeneous fluctuation of temperature $u$ we have:
\begin{eqnarray}
e^{i\phi_h}=-(1+i\pi\frac{1}{n_h}\beta_h u)\nonumber\\
e^{i\phi_l}=-(1+i\pi\frac{1}{n_l}\beta_l u)
\end{eqnarray}

In principle we may take into account here also thermal expansion as $\beta^*_{l,h}=\beta_{l,h}+n_{l,h}\alpha_{l,h}$, however in most of the materials used in multilayer coatings $\alpha\ll\beta$. That is why we omit it here retaining thermal expansion only in thermoelastic noise where the effect is multiplied on the number of layers.

In the first order on $u$:
\begin{eqnarray}
\Gamma_\infty = -1+in_h\pi\frac{\beta_h+\beta_l}{n_l^2-n_h^2} u
\end{eqnarray}

With the cover layer we obtain
\begin{eqnarray}
\Gamma_c&=&\frac{\Gamma_{lh}+\Gamma_\infty}{1+\Gamma_{lh}\Gamma_\infty}e^{i\phi_c}(1+i\pi\frac{4 d_c}{\lambda}\beta_l u)\\
&=&-(1+i\pi n_l u\frac{\beta_h+\beta_l[1- \frac{4dn_l}{\lambda}(1-n_h^2/n_l^2)]}{n_h^2-n_l^2})e^{i\phi_c}\nonumber\\
\Gamma_{out}&=&\Gamma^0_{out}(1+ 4i\pi\beta_{eff} u)\nonumber\\
\Gamma^0_{out}&=&-\frac{n^2_l-1 + (n^2_l+1)\cos\phi_c + 2i n_l\sin\phi_c}{n^2_l+1 + (n^2_l-1)\cos\phi_c}\nonumber\\
\beta_{eff}&=&
\frac{1}{2\pi}\frac{\pi n_l^2(\beta_l+\beta_h)+ \beta_l (\phi_c- \sin\phi_c)(n_h^2-n_l^2)}{(n_h^2-n_l^2)(n_l^2+1+(n_l^2-1)\cos\phi_c)}\nonumber, \label{infinite}
\end{eqnarray}

If $d_c=0$ then we have a QWL multilayer coating starting from high index.
\begin{eqnarray}
\Gamma_{out}=
-(1+i\pi\frac{\beta_h+\beta_l}{n_h^2-n_l^2} u)\nonumber\\
\beta_{eff,H}=\frac{1}{4}\frac{\beta_h+\beta_l}{n_h^2-n_l^2}
\end{eqnarray}

If $d_c=d_l$, then we have QWL multilayer coating starting from low index.

\begin{eqnarray}
\Gamma_{out}=1+i\pi\frac{n_l^2\beta_h+n_h^2\beta_l}{n_h^2-n_l^2} u\nonumber\\
\beta_{eff,l}=\frac{1}{4}\frac{n_l^2\beta_h+n_h^2\beta_l}{n_h^2-n_l^2}
\end{eqnarray}

This expression coincides with the one initially found by BGV in \cite{BGV00} and disproves correction proposed in \cite{BV03}. Recently this expression has been also confirmed by Pinto \cite{Pinto}. Note that the expression for $d_c=0$ could not be simply obtained by indexes $l$ and $h$ reversal. The pitfall here is that these results belong to different branches of solution with $\Gamma_\infty\sim 1$, $Z\to\infty$ for the first low refraction layer and $\Gamma_\infty\sim -1$, $Z\to 0$ for the first high refraction layer.

From the equation (\ref{infinite}) it follows that $\beta_{eff}$ may be considered as a sum of fluctuating term and a term linearly changing with $d_c$. Note that for a  positive variation of temperature and positive $\beta_{l,h}$, phase shift and hence $\beta_{eff}$ is also positive. 

The equation for a a finite number of $2N$ layers may be obtained in the same way as (\ref{finite}) but keeping first order small terms:
\begin{eqnarray}
&&\Gamma_{out,2N} = \nonumber\\ &&-\frac{(n_l^2-1)(1+\varepsilon^2)+(n_l^2+1)(1-\varepsilon^2)\cos\phi_c}{(n_l+\varepsilon)^2+(1+n_l\varepsilon)^2+\cos\phi_c(n_l^2-1)(1-\varepsilon^2)}\nonumber\\
&&+i\frac{2n_l(1-\varepsilon^2)\sin\phi_c}{(n_l+\varepsilon)^2+(1+n_l\varepsilon)^2+\cos\phi_c(n_l^2-1)(1-\varepsilon^2)}\nonumber\\
&&\times\Bigl\{1+\frac{u}{\Lambda}\Bigl[4\varepsilon [N\beta_2n_l/n_h-(N-\cos\phi_c)\beta_l] \nonumber\\
&&+ 2i\pi\frac{\beta^*_l(1-\varepsilon)(n_l^2-\varepsilon n_h^2)+\beta^*_h(1-\varepsilon)^2n_l^2}
{n_h^2-n_l^2}\nonumber\\
&&+ 2i\beta^*_l\phi_c (1-\varepsilon^2)-2i\beta_l(1+\varepsilon^2)\sin\phi_c\Bigr]
\Bigr\}\\
&&\Lambda \equiv (1-\varepsilon^2)(n_l^2+1)+\cos\phi_c(n_l^2-1)(1+\varepsilon^2)\nonumber\\
&&-2i\varepsilon\sin\phi_c(n_l^2-1)
\end{eqnarray}
This equation transforms into (\ref{infinite}) for $\varepsilon\to 0$.

\section*{Appendix B. Photothermal noise in the coating}

To calculate photothermal effect in the coating we apply to BGV model \cite{BGV99}. The power is absorbed in a thin layer $d_r$ of the mirror where the internal intensity decays  $e$ times:
For the QWL coating 
\begin{eqnarray}
d_r=\frac{\lambda(n_l+n_h)}{8n_ln_h\ln(n_h/n_l)}.
\end{eqnarray}
In case of the silica-tantala QWL coating and $\lambda=1.064$ mkm $d_r\simeq 0.43$ mkm. As $d_r\ll w, r_T=a/\sqrt{\omega}$ the model is the following (A.1): 
\begin{eqnarray}
\frac{\partial u}{\partial t}-D\nabla u = 2\frac{w(t)}{\rho C}\frac{2}{\pi w^2} \delta(z)e^{-2(y^2+z^2)/w^2}\nonumber\\
\end{eqnarray}
where $w(t)$ is the fluctuating part of the absorbed power $w(t)=W_{abs}(t)-\langle W_{abs} \rangle$ which for the shot noise has the spectral density $S_{abs}=2\hbar\omega_0 \langle W_{abs}\rangle$. The solution is:
\begin{eqnarray}
u({\bf  r},\omega)= \frac{2w(\omega)}{\rho C} \iiint^{\;\;\;\;+\infty}_{-\infty} \frac{e^{-(k_y^2+k_z^2)w^2/8+i \bf k r}}{Dk^2+i\omega} \frac{d^3k}{(2\pi)^3}
\end{eqnarray}

Temperature fluctuations averaged over the effective volume:
\begin{eqnarray}
\bar u(\omega) &=& \frac{2}{\pi d_r w^2}\iiint^{\;\;\;\;\infty}_{-\infty} e^{-2(y^2+z^2)/w^2-z/d_r} u({\bf r},\omega)\, d^3r \label{ubar}\nonumber\\
&=& \frac{2w(\omega)}{\rho C} \iiint^{\;\;\;\;\infty}_{-\infty} \frac{e^{-w^2(k_y^2+k_z^2)/4}}{(Dk^2+i\omega)(1-ik_xd_r)} \frac{d^3k}{(2\pi)^3} \nonumber\\
&\simeq&  \frac{2w(\omega)}{\rho C} \int^{\infty}_{0}\!\!\int^{\infty}_{-\infty} \frac{e^{-w^2k^2_\bot/4}}{Dk_x^2+i\omega} \frac{k_\bot dk_\bot\, dk_x}{(2\pi)^2} = \nonumber\\ 
&=&\frac{w(\omega)}{\pi w^2\rho C \sqrt{i\omega D}}\sqrt{\pi ib_w}e^{ib_w}[1-{\rm erf}(\sqrt{ib_w})],\nonumber\\
b_w&\equiv& \frac{\omega w^2}{2D}
\end{eqnarray}

Finally as the temperature fluctuations produce phase fluctuation equivalent to $x(\omega) = \beta_{eff}\lambda \bar u(\omega)$ we obtain the spectral density:
\begin{eqnarray}
S^{coat}_{PTR}&=&\frac{S_{abs} (\lambda\beta_{eff})^2}{\pi^2 w^4\rho^2C^2D\omega}G^{coat}_{PT}(\omega),\nonumber\\
G^{coat}_{surf}(\omega)&=&\pi b_w |1-{\rm erf}(\sqrt{ib_w})|^2\nonumber\\
G^{coat}_{surf}(\omega)&\simeq& 1-\frac{5}{4b_w^2}\quad b_w\gg 1\nonumber\\
G^{coat}_{surf}(\omega)&\simeq& \pi b_w-\sqrt{8\pi}b_w^{3/2}\quad b_w\ll 1\nonumber\\
\end{eqnarray}
For very high frequencies when $b_r=\frac{\omega d_r^2}{2D}\gg 1$ this approximation breaks down and should transform into the substrate photothermal noise with substrate parameters substituted by parameters of the coating.

This expression coincides after substituting $S_{abs}$ with that obtained in \cite{Rao} for $b_w\gg 1$.
For practical purposes it is possible to match a simple numerical approximation for $G^{coat}_{surf}$ which has maximum error less 
than 5\% for all $b_w$:
\begin{eqnarray}
G^{coat}_{surf}(\omega)&\simeq& \frac{\pi b_w-2.20b_w^{3/2}+8.98b_w^2}{1+7.44b_w-3.45b_w^{3/2}+8.98b_w^2}
\end{eqnarray}

To calculate the thermoelastic effect of absorbed power in the layer $d_N$ one should at first solve for thermoelastic problem with the given fluctuating temperature to find surface fluctuations which should be averaged over the beam spot.This problem has been solved in \cite{BV03} (A.15) for the Langevin fluctuations:
\begin{eqnarray}
&&\bar x(\omega) =\\
&& 2\alpha d_N (1+\sigma) \iiint^{\;\;\;\;+\infty}_{-\infty} e^{-(k_y^2+k_z^2)w^2/8} u({\bf r},\omega)\, d^3r \nonumber
\end{eqnarray}
Comparing this expression with (\ref{ubar}) we see that boundary conditions on free surface produce more than doubled effect as compared to naive $x(\omega) = \alpha d_N u(\omega)$. Hence we obtain 
\begin{eqnarray}
S^{PTR}_x=\frac{4 S_{abs}(\alpha d_N)^2(1+\sigma)^2}{\pi^2 w^4\rho^2C^2D\omega}G^{coat}_{surf},
\end{eqnarray}
This estimate is two times smaller (for $G^{coat}_{surf}=1$) than that obtained in \cite{Rao} where also Poisson correction is absent, however one should note that in real coating with very inhomogeneous initial temperature distribution over 3-4 layers and  inhomogeneous $\alpha$ and $\rho C$, coefficients may sufficiently differ from these rough approximations and from averaging adapted from surface thermoelastic calculations in \cite{Fejer} high frequency limit of this approximation is reached at significantly lower frequencies, when
$b_d=\frac{\omega d_N^2}{2D}\gg 1$, where the effect is transformed into the bulk photothermal effect of BGV \cite{BGV99} in the effective layer material.

\section*{Appendix C. Thermorefractive and thermoelastic noises}

The calculation of cross-correlation effects could not be easily obtained using simple fluctuation-dissipation-theorem approach. In this case the Langevin fluctuating forces are valuable. Using the previous BGV results the spectrum of the fluctuations of temperature in the mirror may be found as:
\begin{eqnarray}
&&u(\omega,{\bf r})= \iiint^{\;\;\;\;+\infty}_{-\infty}
        \frac{F(\vec k,\omega) e^{i{\bf k r}}}{D k^2+ i\omega}\,\frac{d^3k}{(2\pi)^3}\\
&&\langle\,F({\bf k},\omega)F^*({\bf k'},\omega')\, \rangle=\nonumber\\
&&(2\pi)^4\,
\frac{2\kappa T D}{\rho C} k^2 \delta({\bf k}-{\bf k'})\, \delta (\omega-\omega')\nonumber
\end{eqnarray}
Thermorefractive fluctuations in a very thin effective layer $d_r$ where the reflection mostly occurs are proportional to the exponent-weighted averaging over the effective reflective layers $d_r$.
\begin{eqnarray}
&&\bar X^{coat}_{TR}(\omega)=\\
&&-\beta_{eff}\lambda \frac{2}{\pi d_r w^2} \iint^{\;\;\;\;+\infty}_{-\infty}\int^{+\infty}_{0}e^{-x/d_r-2(y^2+z^2)/w^2}u(\omega,{\bf r})\nonumber\\
&&=-\beta_{eff}\lambda  \iiint^{\;\;\;\;+\infty}_{-\infty}\frac{F_s(\vec k,\omega) e^{-k^2_\bot w^2/8}}{(D_s k^2+ i\omega)(1-ik_x d_r)}\frac{d^3k}{(2\pi)^3}, \nonumber
\end{eqnarray}
where $k_\bot^2=k_y^2+k_z^2$. As the layer is thin the main contribution to the fluctuations of temperature in the layer is produced by substrate. As it is shown in \cite{Fejer} the contribution of the layer itself is $\sqrt{\omega d_N^2/D_c}$ times smaller. That is why all the parameters under integral are related to the substrate. The effect of the layer becomes apparent due to thermal expansion and fluctuation of the refraction index.

The calculation of the effect produced by thermoelastic effect in the whole multilayer coating on the substrate is a more difficult problem. It requires the solution of the thermal and elasticity problems with a subsequent averaging of the surface displacement. However as the spectral density has been already accurately calculated \cite{BV03,Fejer}, one may in the first order on $d_N$ use analogous averaging with effective thickness $d^*\sim d_N$ fitted to the known results.
An accurate straightforward calculation for inhomogeneously distributed noise sources may be performed using the approach of Green's functions for the Langevin sources developed by van Vliet \cite{Vlietfilm, vanVliet}.
\begin{eqnarray}
&&\bar X^{coat}_{TE}(t)\simeq\\
&& \frac{2\alpha_c d_N (1+\sigma_s)}{(2\pi)^3}\,
        \iiint^{\;\;\;\;+\infty}_{-\infty}
        \frac{F_s({\bf k},\omega)e^{-k^2_\bot w^2/8}\,d^3k}{(D_s k^2+ i\omega)(1-ik_xd^*)}\nonumber
\end{eqnarray}
We obtain:
\begin{eqnarray}
&&S^{coat}_{T}(\omega)=S^{coat}_{TE}+S^{coat}_{TR}-S^{coat}_{TER} =\\
&&\frac{8D_sk_B T^2}{(2\pi)^3\rho_s C_s}\iiint^{\;\;\;\;+\infty}_{-\infty}\left|\frac{2\alpha_cd_N(1+\sigma_s)}{1-ik_xd^*}-\frac{\beta_{eff}\lambda}{1-ik_xd_r}\right|^2\nonumber\\
&&\times\frac{k^2 e^{-k^2_\bot w^2/4}d^3k}{D_s^2 (k_\bot^2+k_x^2)^2+ \omega^2}\nonumber
\end{eqnarray}

\begin{eqnarray}
&&S^{coat}_{TE}\simeq\frac{32D_sk_B T^2 \alpha_cf^2d_N^2(1+\sigma_s)^2}{(2\pi)^3\rho_s C_s}\nonumber\\
&&\times\iiint^{\;\;\;\;+\infty}_{-\infty}\frac{1}{1+k_x^2d^{*2}}\frac{k^2 e^{-k^2_\bot w^2/4}}{D_s^2 k_x^4+ \omega^2}d^3k\nonumber\\
&&=\frac{8k_B T^2 \alpha_c^2d_N^2(1+\sigma_s)^2}{\pi^{3/2} w^2 \sqrt{\kappa_s \rho_s C_s} f^{1/2}}
\frac{1-\sqrt{2}b+b^2}{1+b^4},\\
&&b=\sqrt{\omega d^{*2}/D_s} \nonumber
\end{eqnarray}

From (\ref{gappr}) it follows that $\sqrt{2}b\simeq \frac{3R^2-1}{3R}\xi$ and hence $d^*\simeq d_N  \frac{3R^2-1}{3R}$.

\begin{eqnarray}
&&S^{coat}_{TR}\simeq \frac{2k_B T^2 \beta_{eff}^2\lambda}{\pi^{3/2} w^2 \sqrt{\kappa \rho C} f^{1/2}}
\frac{1-\sqrt{2}c+c^2}{1+c^4},\\
&&c=\sqrt{\omega d_r^2/D} \nonumber\\
&&S^{coat}_{TER}\simeq\nonumber\\
&&\frac{8k_B T^2 \alpha_cd_N(1+\sigma_s)\beta_{eff}\lambda}{\pi^{3/2}\sqrt{\kappa \rho C} w^2 f^{1/2}}
\frac{1-\sqrt{2}(b-c)+b(b-c)}{1+b^4}\nonumber
\end{eqnarray}

We may now choose a frequency of interest $f_{opt}$ and optimize for it the combined thermoelastic and thermorefractive noise. Considering $c\ll b$:
\begin{eqnarray}
\beta_{eff,opt}\lambda &\simeq& 2\alpha_c d_N(1+\sigma_s) B(f_{opt})\nonumber\\
S^{coat}_{T,opt} &\simeq& S^{coat}_{TE} B(f)[1-B(f)]+[B(f_{opt})-B(f)]^2\nonumber\\
B(f)&\equiv&\frac{1-\sqrt{2}b+b^2}{1+b^4}\nonumber\\
\end{eqnarray}

In the calculations I assumed that in the area of interest $d_r\ll d_N\ll w$ and $k_\bot \ll k_x$. The approximation can possibly break down when $\xi \sim 1$ which is attained in silica-tantala coating with $~20$ double layers at $f\sim 1 kHz$. In this case thermal sources in the coating itself should be additionally considered. 

\section*{Appendix D. Substrate and coating thermoelastic noise correlation}

The effects of thermoelastic substrate and coating effects are the following:
\begin{eqnarray}
&&\bar X^{sub}_{TE}(\omega)= \nonumber\\
&&2\alpha_s (1+\sigma_s)\,
        \iiint\limits_{-\infty}^{\;\;\;\;\;\infty}
    \frac{d^3k}{(2\pi)^3}\,
        \frac{F({\bf k},\omega)}{D_s k^2+ i\omega}\,
        \frac{k_\bot}{k^2 }\,
    e^{i\omega t-k^2_\bot w^2/8},\nonumber\\
&&\bar X^{coat}_T(\omega)=-\beta_{eff}\lambda \bar u =\nonumber\\
&&-\beta_{eff}\lambda \iiint\limits_{-\infty}^{\;\;\;\;\;\infty}\frac{d^3 k}{(2\pi)^3}\
    \frac{F({\bf k},\omega)}{D_s k^2 + i\omega}\
    e^{-k^2_\bot w^2/8},\nonumber\\
&& S_{\Sigma}= \frac{8D_sk_B T^2}{(2\pi)^3\rho_s C_s}
\iiint\limits_{-\infty}^{\;\;\;\;\;\infty}\left[2\alpha_s(1+\sigma_s)\frac{k_\bot}{k^2 }-\beta_{eff}\lambda\right]^2\times\nonumber\\ 
&&\frac{k^2 e^{-k^2_\bot w^2/4}}{D_s^2 (k_\bot^2+k_x^2)^2+ \omega^2}d^3k\nonumber\\
&&=S^{coat}_{T}+S^{sub}_{TE}-S_{TER},\nonumber\\
 &&S^{sub}_{TE}\simeq    
    \frac{4 k_B T^2 \alpha_s^2(1+\sigma_s)^2\kappa_s}{\pi^{5/2}(\rho_s C_s)^2 w^3f^2} \nonumber\\
&&S^{coat}_{T}\simeq\frac {2k_B T^2\beta_{eff}^2\lambda^2}{\pi^{3/2}\,w^2 \sqrt{\kappa_s\rho_s C_s}f^{1/2}}
\nonumber\\
 &&S_{TER}=\frac{8 D_sk_B T^2}{\pi^2\rho C}\alpha_s (1+\sigma_s)\beta_{eff}\lambda \nonumber\\
 &&\times \iint\limits_{-\infty}^{\infty}e^{-k^2_\bot r_0^2/2}
        \frac{k_\bot^2 dk_\bot dk_x}{D_s^2 k^4+ \omega^2}\, \nonumber\\
  &&\simeq\frac{8 k_B T^2 \sqrt{\kappa_s}\alpha_s (1+\sigma_s)\beta_{eff}\lambda}{\pi^2 (\rho_s C_s)^{3/2}f^{3/2}w^3}   
\end{eqnarray}

Now optimizing $\beta_{eff}$ at a frequency of interest we find:
\begin{eqnarray}
  S_{\Sigma, min}&=&S^{sub}_{TE}\left(1-2\sqrt{\frac{D_s}{w^2f}}\right)
\end{eqnarray}

\section*{Appendix E. Parameters used in numerical estimates \cite{Fejer}}

General parameters:
\begin{eqnarray}
\begin{array}{ll}
T=300\,\mbox{K}&\lambda=1.064\times 10^{-6}\,\mbox{m}\nonumber\\
\omega_0=1.77\times 10^{15}\,\mbox{s}^{-1} &w=0.06\mbox{m}\nonumber\\
\end{array}
\end{eqnarray}
Fused silica substrate parameters:
\begin{eqnarray}
\begin{array}{lll}
&\alpha_s=5.1\times10^{-7}\ \mbox{K}^{-1}, &\kappa_s=1.38\, \mbox{J/(m s K)}, \label{parameter}  \nonumber\\
&\rho=2200\, \mbox{kg/m}^3,  &C=746\,\mbox{J/(kg K)}, \nonumber\\
&Y_s=7.2\times 10^{10\,}\mbox{N/m}^2, &\sigma=0.17, \nonumber\\
&\phi=5\times10^{-9}
\end{array}
\end{eqnarray}
Thin film parameters (Silica and tantala):
\begin{eqnarray}
\begin{array}{llll}    
n_l=1.45 &\phi_l=1.0\times10^{-4}\,\nonumber\\
\alpha_l=\alpha_s & \beta_l=1.5\times10^{-5} \mbox{K}^{-1} \nonumber\\
\kappa_l=\kappa_s & \rho_l=\rho_s, \nonumber\\
n_h=2.065&\phi_h=3.8\times10^{-4}\nonumber\\
\alpha_h=3.6\times10^{-6}\ \mbox{K}^{-1}, &\beta_l=6\times10^{-5}\, \mbox{K}^{-1}, \nonumber\\
\kappa_h=33\, \mbox{J/(m s K)}, &\rho_h=6850\,\mbox{kg/m}^3\nonumber
\end{array}
\end{eqnarray}

\end{document}